\documentclass[aps,amsfonts,prb,twocolumn,showpacs ,superscriptaddress]{revtex4-1}

 \usepackage{setspace}
 \usepackage{color}

\usepackage{amsmath}
\usepackage{amsfonts}            
\usepackage{amssymb}

\usepackage{mathrsfs}

\usepackage{eufrak}

\usepackage{graphicx}
\usepackage{epstopdf}

\usepackage{dcolumn}
\usepackage{bm}

\makeatletter
\def\@dotsep{4.5}
\makeatother

\def\ba{\begin{eqnarray}}
\def\ea{\end{eqnarray}}
\def\beq{\begin{equation}}
\def\eeq{\end{equation}}

\begin{document}


\title{Controllable quantum spin glasses with magnetic impurities \\ embedded in quantum solids}

\author{Mikhail Lemeshko} 
\email{mlemeshko@cfa.harvard.edu}
\affiliation{ITAMP, Harvard-Smithsonian Center for Astrophysics, 60 Garden Street, Cambridge, MA 02138, USA}%
\affiliation{Physics Department, Harvard University, 17 Oxford Street, Cambridge, MA 02138, USA} %
\affiliation{Kavli Institute for Theoretical Physics, University of California, Santa Barbara, CA 93106, USA}

\author{Norman Y. Yao}
\affiliation{Physics Department, Harvard University, 17 Oxford Street, Cambridge, MA 02138, USA} %

\author{Alexey V. Gorshkov}
\affiliation{Institute for Quantum Information and Matter, California Institute of Technology, Pasadena, CA 91125, USA}
\affiliation{Kavli Institute for Theoretical Physics, University of California, Santa Barbara, CA 93106, USA}

\author{Hendrik~Weimer}
\affiliation{Institut f\"ur Theoretische Physik, Leibniz Universit\"at Hannover, Appelstr. 2, 30167 Hannover, Germany}

\author{Steven D. Bennett}
\affiliation{Physics Department, Harvard University, 17 Oxford Street, Cambridge, MA 02138, USA} %

\author{Takamasa Momose}
\affiliation{Department of Chemistry, The University of British Columbia, Vancouver, BC V6T 1Z1, Canada}

\author{Sarang Gopalakrishnan}
\affiliation{Physics Department, Harvard University, 17 Oxford Street, Cambridge, MA 02138, USA} %

\begin{abstract}
Magnetic impurities embedded in inert solids can exhibit long coherence times and interact with one another via their intrinsic anisotropic dipolar interaction. We argue that, as a consequence of these properties, disordered ensembles of magnetic impurities provide an effective platform for realizing a controllable, tunable version of the  dipolar quantum spin glass seen in  LiHo$_x$Y$_{1-x}$F$_4$. Specifically, we propose and analyze a system composed of dysprosium atoms embedded in solid helium. We describe the phase diagram of the system and discuss the realizability and detectability of the quantum spin glass and antiglass phases. 
\end{abstract}


\date{\today}

\maketitle


\section{Introduction}

Dipolar interactions between spins give rise to a wealth of exotic phases in condensed-matter systems, of which one of the most striking is the possible quantum spin glass in lithium holmium fluoride (LiHo$_x$Y$_{1-x}$F$_4$)~\cite{rosenbaum87, ReichPRB90, rosenbaum:qsg, asg2002, kycia2008, muonspin2010}. Equilibrium measurements on this material indicate the presence of a low-temperature spin glass phase for holmium concentrations, $x \alt 0.25$. However, numerous surprising properties of a spin glass, especially those related to its quantum-coherent far-from-equilibrium dynamics, are difficult to directly probe in solid-state systems,  due to the impact of decoherence and relaxation channels present. Moreover, a number of questions concerning the equilibrium phase diagram are still open, e.g.\ whether a low-temperature spin liquid or ``antiglass'' phase exists for $x \leq 0.05$, and what are the consequences of the transverse field on the phase diagram~\cite{qsgreview, asg2002}. 

The above questions are challenging to address in the context of lithium holmium fluoride; for example, applying a transverse magnetic field  induces random longitudinal fields as a side-effect~\cite{qsgreview}. 
 This motivates the realization of dipolar quantum spin glasses in a well-isolated setting with long coherence times. Although there have been proposals to mimic the properties of spin and charge glasses using ultracold atoms coupled to multimode cavities~\cite{sg2009, sg2011, StrackPRL11, strack2012}, it is also desirable to naturally realize a spin glass using only  bare dipolar interactions. Such a realization would amount to a ``quantum simulator'' and could shed light on the dynamical properties of quantum glasses~\cite{asg2002}. 
 Generically, however, such a realization is challenging for two related reasons. First, the number of ultracold atoms achievable in experiment is inherently \emph{small} relative to condensed matter systems;  this limitation is particularly severe when studying the physics of spin glasses, since the properties of disordered systems are dominated by isolated rare events, suppressed in small systems~\cite{griffiths1969, young:griffiths}. Second, the dipole-dipole interaction corresponding to experimentally feasible atomic densities is typically weak, rendering the interaction timescale comparable to the decoherence timescale. 
These difficulties motivate us to consider \emph{scalable} realizations of dipolar spin models in which the dipoles can be brought close to one another while maintaining long coherence times.

In this work we discuss the feasibility of a scalable platform based on ``matrix isolation,'' i.e.  atoms or molecules embedded in inert matrices such as solid  helium, parahydrogen, or rare gases~\cite{WhittleJCP54}. The technique of matrix isolation is frequently used in chemical physics for high-resolution spectroscopy of individual molecules and to study basic chemical reactions. In addition, this technique allows  one to acquire the  spectra of chemical species, such as radicals, whose reaction times are extremely  fast in the gas phase~\cite{MomoseBCSJ98, MomoseVibSpec04, MomoseIRPC05}. For sufficiently inert matrices, the spectral lines of matrix-isolated species are sharper than those in the gas phase, owing to the absence of motional broadening---the atoms can be regarded as both fixed in space and undisturbed by their environments.  Under these conditions, individual atoms possess very long coherence times ($T_1$, $T_2$), and can be optically pumped, with high efficiency, into a given internal state or set of internal states~\cite{KanorskyPRA96}. As we show below, a matrix-embedded atom can be regarded as a highly controllable, quantum-coherent degree of freedom featuring strong dipolar interactions, analogous to ultracold atoms in optical lattices~\cite{BlochRMP08} and solid-state defect centers~\cite{JelezkoNJP12, KoehlNat11}.

While the distances between atoms or molecules in optical lattices are limited by the optical wavelength of the trapping laser to a minimum of a few hundred nanometers,    the separations between  matrix-isolated atoms can be in principle as small as a few tens of nanometers~\cite{MomoseVibSpec04}. It is worth noting that  recently proposed nanoplasmonic structures can potentially allow to trap atoms at subwavelength distances~\cite{GullansPRL12}, however, these  do not allow to create an ensemble of atoms randomly distributed  in three dimensions. 
The coherence times for matrix-isolated species can be on the order of 1 second~\cite{KanorskyPRA96} at temperatures of about 1~K; atoms do not need to be cooled down to nanoKelvin temperatures as it is the case in optical lattice experiments. Furthermore, the range of atoms and molecules that can be matrix isolated is quite broad and is not limited to the species that can be laser cooled~\cite{MetcalfBook, ShumanNature10} or associated from ultracold atoms~\cite{KohlerRMP06, JonesRMP06}. In a way, atoms or molecules trapped in matrices behave similarly to defects in solids~\cite{JelezkoNJP12, KoehlNat11}. However, such impurities can feature substantially higher magnetic or electric dipole moments compared to defect centers, resulting in significantly enhanced interaction strengths. 

To be specific, in this work we consider dysprosium atoms embedded at random sites of a solid helium matrix, and discuss the possibility of realizing a dipolar quantum spin glass in such a system. Dysprosium atoms strongly interact  with one another, but not, as we shall argue, with the inert matrix. We indicate a scheme whereby the dysprosium atoms can be optically pumped into a combination of two magnetic states, which collectively form an Ising spin. Thus, an interacting system of many dysprosium atoms can be mapped onto an Ising model with dipolar interactions. Once randomness is included, the effective model is precisely the one studied in Refs.~\cite{rosenbaum:qsg, young:griffiths}, which is believed to describe the physics of LiHoF.

The paper is organized as follows. In Sec.~\ref{sec:impurities} we discuss the relevant energy scales for magnetic impurity atoms in inert matrices, and present the optical pumping scheme that can be used to realize Ising spins. In Sec.~\ref{sec:spin-glass} we describe the expected phase diagram and provide a recipe to detect and control the achievable phases, including the spin glass and antiglass. In Sec.~\ref{sec:magnetism} we discuss possible applications of the platform to exotic quantum magnetism. We compare the matrix isolation technique to other ways of achieving strongly interacting dipole ensembles and summarize the conclusions of the present work in Sec.~\ref{sec:conclusions}.

\section{Magnetic atom impurities in inert matrices}
\label{sec:impurities}

In this section we provide the estimates for coherence times and optical pumping efficiency in a system of matrix-isolated magnetic atoms, based on previously obtained experimental data. 
As an impurity species we choose dysprosium -- the most magnetic atom in the periodic table. Its bosonic and fermionic isotopes have been recently brought to  quantum degeneracy~\cite{LuPRL11, LuPRL12}, which provides us a benchmark for comparison with other ultracold atomic experiments. We note that using matrix-isolated polar molecules~\cite{FajardoJCP09} can potentially result in stronger dipole-dipole interactions between the impurities,  however, decoherence due to coupling to phonons is also stronger for species possessing an electric dipole moment~\cite{ArndtPRL95, KatsukiMomosePRL00}.

Light matrices, such as solid \textit{para}-H$_2$ and solid helium are softer and less polarizable compared to heavier ones (solid Ne, Ar, Xe, etc.)~\cite{MomoseVibSpec04, MoroshkinPhysRep08}. This allows to trap atoms leaving their energy states almost intact and to achieve long coherence times due to weak interaction with phonons. Solid hydrogen is largely inert, and does not react with atoms such as I or Cl; however, there have been observed reactions of H$_2$ with Ba and Hg in the excited electronic state~\footnote{T. Momose, unpublished}, which might complicate optical pumping. Since the reactions of H$_2$ with magnetic rare-earth atoms are largely unexplored, in this work we focus on solid helium as a matrix that has been proven to be nonreactive.  The helium matrix behaves as an amorphous solid close to a liquid, therefore the crystal fields and phonon coupling strengths  are typically weak~\cite{UlzegaThesis, MoroshkinPhysRep08}. This motivated a number of studies of atomic impurities in He to measure the electron electric dipole moment and to do precision spectroscopy~\cite{KozlovPRL06, UlzegaThesis, MoroshkinPhysRep08}. Kanorsky~\textit{et al.} demonstrated optical pumping of cesium atoms trapped in solid He and performed  ESR spectroscopy with milliHertz resolution~\cite{KanorskyPRA96}. The group of Weis performed detailed optical spectroscopic studies of Cs atoms in solid He~\cite{HoferPRA07, MoroshkinPRA13}. Magnetic rare-earth atoms such as thulium have also been isolated and studied in solid He, and narrow lines for electronic transitions have been observed~\cite{IshikawaPRB97}. Gordon~\textit{et al.} demonstrated another technique allowing to achieve high impurity densities in solid Helium, corresponding to interparticle distances of a few tens of nm.\cite{GordonJLTP03, GordonJLTP05}

The ESR coherence times, $T_1$ and $T_2$, for Cs atoms in solid He have been measured to be on the order of $1-2$ s and $0.1-0.2$ s respectively~\cite{KanorskyPRA96, MoroshkinLTPhys06}. Here we discuss two main sources of decoherence in matrix-isolated dysprosium: (i) spin-phonon relaxation, and (ii) dephasing due to the nuclear spin bath of  $^3$He impurities. 


(i) We estimate the spin-phonon relaxation rates within the most general framework. We consider the Hamiltonian $H = H_0 + V_\text{s-ph}$, where 
\begin{equation}
\label{H0eq}
	H_0 = \frac{\Delta}{2} \sigma_z + \sum_k \epsilon_k a_k^{\dagger} a_k
\end{equation}
corresponds to the uncoupled part and
\begin{equation}
\label{phonon1}
	V_\text{s-ph} = \sigma_x \sum_k \lambda_{kx} (a_{kx} + a_{kx}^{\dagger} ) + \sigma_z \sum_k \lambda_{kz} (a_{kz} + a_{kz}^{\dagger}).
\end{equation}
is the spin-phonon interaction. Here $\sigma_{x,z}$ are Pauli operators describing the impurity atom as an effective spin-$1/2$ system, and $a_{k}^{\dagger}$, $a_{k}$ give the creation and annihilation operators for a phonon in mode $k$.
We allow for the possibility that the spin couples to different phonon operators (indexed with $x$ and $z$) in the longitudinal and transverse directions, with corresponding strengths $\lambda_{kx}$ and $\lambda_{kz}$. At linear order in the spin-phonon interaction, the influence of the phonons is determined purely by the spectral density,
\begin{equation}
 \label{phonon2}
	J_{\xi} (\omega) = \frac{\pi}{2} \sum_k \lambda_{k\xi}^2 \delta (\omega - \omega_k)
\end{equation}
where $\xi = x,z$. The $\sigma_x$ term of eq.~(\ref{phonon1}) allows for the direct absorption or emission of a phonon with energy $\varepsilon$ accompanied by a spin-flip. Using Fermi's golden rule, the total rate for such a direct process is:
\begin{equation}
 \label{phonon3}
	\frac{1}{T_1^D}=  4 J(\varepsilon) [2n(\varepsilon) +1 ],
\end{equation}
where $n(\varepsilon) = 1/(e^{\beta \varepsilon} -1)$ is the boson occupation number. As expected,  direct phonon-induced spin-flip processes are important only at very low temperatures, because for $\varepsilon \ll T$ the relevant density of states is highly suppressed.  Thus, typically, one finds that two-phonon Raman-type processes~\cite{JarmolaPRL12} dominate the phonon-induced relaxation since they make use of the entire phonon spectrum. From Fermi's golden rule at second order in coupling,
\begin{align}
 \label{phonon4}
	\frac{1}{T_1^R}&= \frac{32}{\pi} \int d \omega [2n(\omega) n(\omega +\varepsilon) + n(\omega) + n(\omega +\varepsilon) ] \nonumber \\
	 &\left [ \frac{J_x (\omega +\varepsilon) J_z (\omega)}{\omega^2} + \frac{J_x (\omega ) J_z (\omega+\varepsilon)}{(\omega+\varepsilon)^2}  \right ].
\end{align}
In solid He, we expect the relaxation to be dominated by  acoustic phonons whose density of states scales as $\sim \omega^2$; since the matrix elements scale as $\sqrt{\omega}$, one finds that $J \sim \omega^3$. Such an approach is consistent with phenomenological decay associated with nitrogen-vacancy centers in diamond~\cite{JarmolaPRL12}. While further experimental input is necessary to provide the spectral density, one can make a conservative estimate for the $T_1$ time of Dy based on experiments performed with matrix-isolated Cs~\cite{KanorskyPRA96}. Assuming that the larger magnetic dipole moment of Dy enhances its coupling to phonons (a fact which depends on the microscopic coupling mechanism), we expect, at worst, $T_1 \sim 10-100$~ms.

(ii)  Now we consider dephasing due to the nuclear spin bath. Since the concentration of $^3$He impurities which carry a nonzero nuclear spin is only 0.000138\%, this will likely not be the dominant source of dephasing. Based on the experimental results for Cs~\cite{KanorskyPRA96, MoroshkinLTPhys06}, we use a simple scaling argument in order to account for a larger magnetic dipole moment of Dy. The resulting coherence time $T_2$ due to the nuclear spin bath is approximately 100~ms.

\begin{figure}[t]
\includegraphics[width=0.95\linewidth]{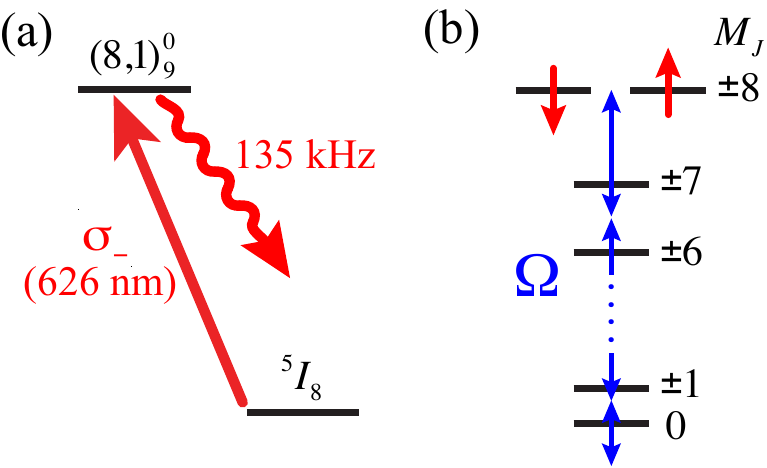}
\caption{(a) Dysprosium atoms are optically pumped into the $M_J=-8$ component of the ground $^5 I_8$ electronic state, using $\sigma_-$-polarized 626 nm excitation to the $(8,1)^o_9$ state. (b) The ground $J=8$ level is split into 9 $|M_J|$ components with a $\pi$-polarized light far detuned from the 626 nm transition. A multi-photon transverse field $\Omega$ connects the effective spin states, $M_J = -8$ and $M_J = +8$, without populating any intermediate levels.}
\label{levels}
\end{figure}

Let us consider a bosonic isotope of dysprosium, e.g.\ $^{162}$Dy or $^{164}$Dy, whose ground electronic state is $^5 I_8$ (in the $^{2S+1}L_J$ notation) possessing a dipole moment of $10\mu_B$ due to the spin and orbital electronic degrees of freedom. As the first step, we initialize the system by optically pumping all the Dy atoms into the $M_J=-8$ component using $\sigma_-$-polarized 626~nm excitation to the $(8,1)^o_9$ state (with a linewidth of $\gamma = 135$~kHz)~\cite{LuPRA11}, cf.~Fig.~\ref{levels}(a). Assuming that relaxation redistributes the population randomly between all the $M_J$ states at rate corresponding to $T_1 =100$~ms, and the regime of the pumping Rabi frequency $\Omega_\text{pump} \gg \gamma$, we estimate the optical pumping efficiency to be 94\%. We note that the many-body effects described in the following sections should be detectable provided that the pumping efficiency significantly exceeds 50\%.

Second, we use a $\pi$-polarized laser field~\cite{SmithPRL04} far detuned from the 626 nm transition, in order to achieve an anharmonic splitting of the $J=8$ level into nine $|M_J|$ components, cf.~Fig.~\ref{levels}(b).  A transverse field $\Omega$ is realized by a $16$-photon transition coupling $M_J = -8$ and $M_J = +8$, by analogy to recent experiments on alkali atoms~\cite{ChaudhuryPRL07, MerkelPRA08} and diatomic molecules~\cite{OspekausPRL10}. Due to the anharmonicity of the level structure, using at most $8$ microwave fields  polarized along $x$ or $y$ and far-detuned from the intermediate levels allows to couple the $M_J = \pm 8$ states directly, without populating any other states. Furthermore, the anharmonicity of the level structure allows to avoid the `flip-flop' processes driving the atoms out of the spin-$1/2$ manifold.

\section{Spin-glass physics}
\label{sec:spin-glass}

\begin{figure}[tl]
\begin{center}
\includegraphics[width=\linewidth]{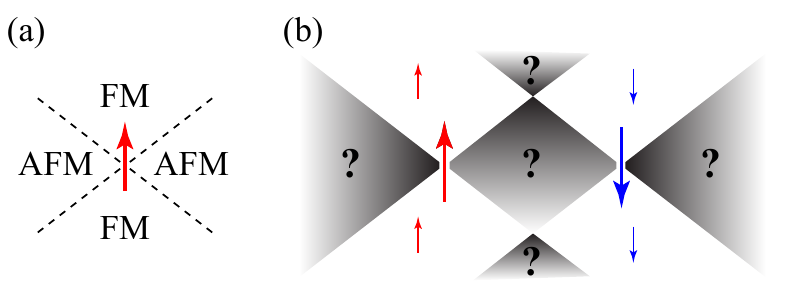}
\caption{(a) Interaction of a test spin with a single dipolar spin, indicating ferromagnetic (FM) and antiferromagnetic (AFM) regions. (b) Interaction of a test spin with two dipolar spins. Shaded regions indicate where the interaction is frustrated.}
\label{frustration}
\end{center}
\end{figure}

In this section we describe the many-body model that we propose to engineer  with magnetic atoms randomly distributed inside of an inert matrix. Denoting the states $M_J = - 8$ and $M_J = 8$ as $\vert \! \downarrow \rangle$ and $\vert \! \uparrow \rangle$, respectively, the effective Hamiltonian of the system is that of a transverse-field Ising model with dipolar interactions:
\begin{equation}
\label{Ising}
	H=  \sum_i (\Omega \sigma^{i}_x + \Delta \sigma^{i}_z) +  \frac{V_\text{dd}}{2} \sum_{i \neq j} \frac{1 -3 \cos^2 \theta_{ij}}{\vert \mathbf{r}_i - \mathbf{r}_j\vert^3} \sigma^{i}_z \sigma^{j}_z,
\end{equation}
where $V_\text{dd} = \mu_0 \mu^2/(4\pi)$, with $\mu_0$ the vacuum permeability and $\mu$ the magnetic dipole moment of impurity atoms ($\mu = 10 \mu_B$ in case of Dy). The quantization axis of the spins is determined by the linear polarization of the off-resonant laser, used to achieve the anharmonic splitting of the $J=8$ level, as described in Sec.~\ref{sec:impurities} (this laser stays on throughout the experiment). The detuning $\Delta$ of the microwave field will be used for preparing the desired ground states of the transverse-field Ising model. 
 The vectors $\mathbf{r}_i$ and $\mathbf{r}_j$ give the positions of the interacting dipoles, with $\mathbf{r}_i - \mathbf{r_j} \equiv \mathbf{R}_{ij} = (R_{ij},\theta_{ij},\phi_{ij})$. 
The dipole-dipole interaction of Eq.~(\ref{Ising}) is evidently ferromagnetic for spins aligned head-to-tail, $\theta_{ij}=0$, and antiferromagnetic for spins aligned side by side, $\theta_{ij} = \pi/2$.  As shown in Fig.~\ref{frustration}, this interaction is in general frustrated, therefore it is plausible that a system of randomly arranged dipoles can possess a spin glass phase.

\begin{figure}[tr]
\begin{center}
\includegraphics[width=0.8\linewidth]{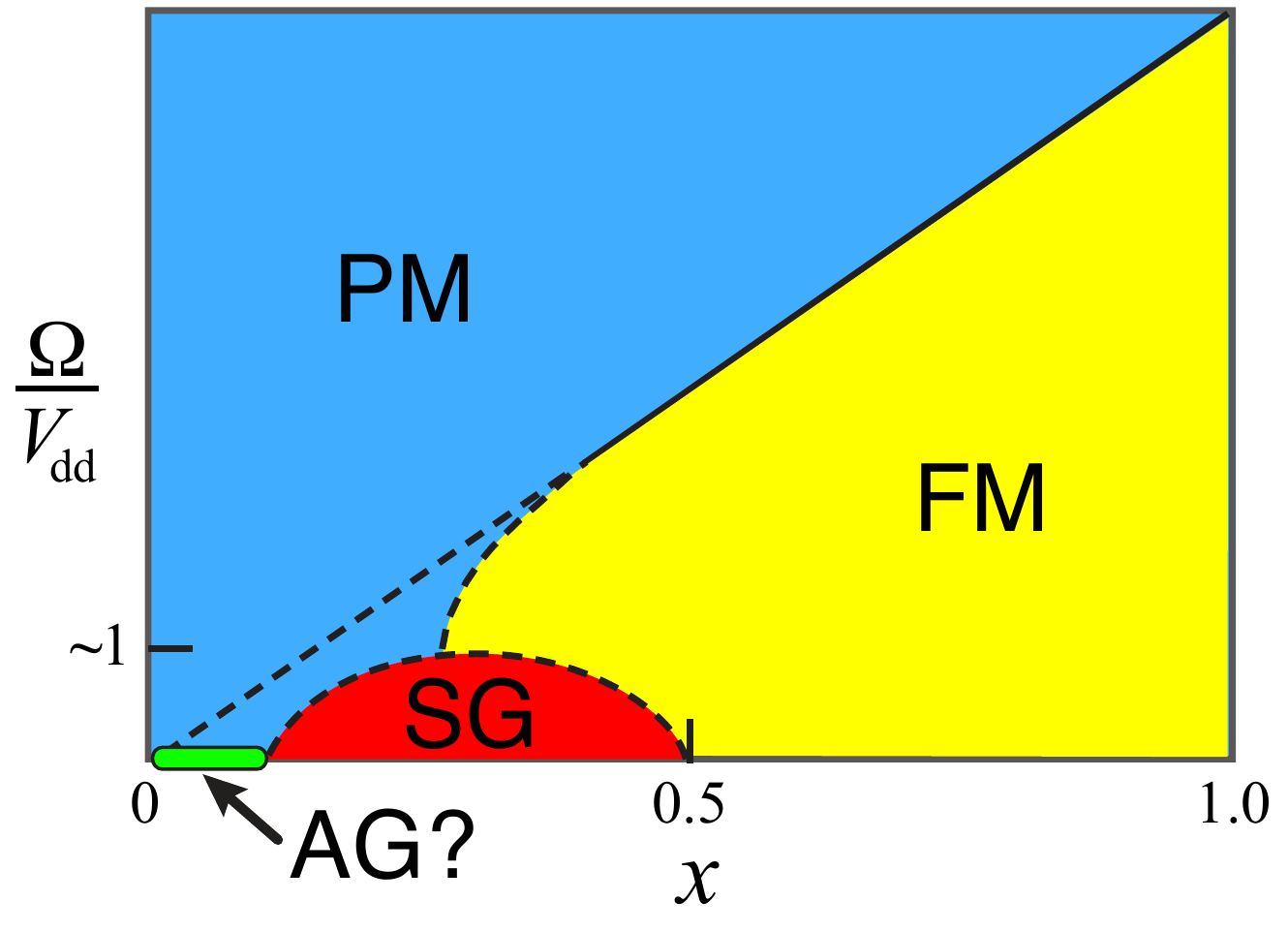}
\caption{Phase diagram of the system as a function of spin concentration $x$, featuring paramagnetic (PM), ferromagnetic (FM), spin-glass (SG), and a possible antiglass (AG) phases, see Refs.~\cite{ReichPRB90,qsgreview}}
\label{phase}
\end{center}
\end{figure}

The Hamiltonian (\ref{Ising}) naturally occurs in lithium holmium fluoride (LiHo$_x$Y$_{1-x}$F$_4$)~\cite{ReichPRB90}.  In this material a fraction $x$ of the lattice sites are randomly substituted by Ising spins (Ho) which interact via a dipolar interaction, resulting in the phase diagram shown in Fig.~\ref{phase}. If every lattice site is occupied by an Ising spin, the ground state is ferromagnetic, however a spin glass phase is known to occur for a wide range of intermediate concentrations. Finally, as a result of quantum fluctuations, there appears to be an ``antiglass'' phase~\cite{ReichPRB90} as $x \rightarrow 0$, cf. Fig.~\ref{phase}, whose existence is still under active discussion~\cite{qsgreview}.
In a system of matrix-isolated magnetic atoms, instead of temperature, we use a tunable transverse field $\Omega$ to destabilize the ordered states. In the Ising model, eq.~(\ref{Ising}), a transverse field and a temperature typically have similar effects on the phase diagram~\cite{qsgreview, SachdevQPT}, i.e., they both destabilize magnetically ordered states.



We now discuss how the various phases of Fig.~\ref{phase}, including the spin-glass, can be initialized and explored using matrix-isolated magnetic atoms. We start by optically pumping all the atoms into state $\vert \! \downarrow \rangle$, in the absence of microwave fields. Then we apply a microwave field with a large detuning, $\Delta \gg \Omega, V_{dd}$, so that the  state with all atoms in $\vert \! \downarrow \rangle$ is the ground state of the Hamiltonian, Eq.~(\ref{Ising}). The ferromagnetic phase can be prepared by properly choosing the spin concentration $x$ and the ratio $\Omega/V_{dd}$, cf. Fig.~\ref{phase}, and adiabatically turning off the detuning $\Delta$. The extent of the phase can be explored by adiabatically tuning $\Omega$. 

In order to prepare the paramagnetic phase, after the optical pumping, we apply a microwave field satisfying $\Delta \gg \Omega \gg V_{dd}$. Then we turn off the detuning $\Delta$ adiabatically relative to $\Omega$ but diabatically relative to $V_{dd}$, which prepares each atom in state $\vert \leftarrow \rangle = (\vert \uparrow \rangle - \vert \downarrow \rangle)/\sqrt{2}$. Since $\Omega \gg V_{dd}$, this prepares the paramagnetic ground state of the Hamiltonian. Adiabatically tuning $\Omega$ allows one to explore the extent of the paramagnetic phase. It is worth pointing out that, equivalently, one could prepare all atoms in state $\vert \leftarrow \rangle$ by applying a fast $\pi/2$ pulse around the $\hat y$ axis and then turning on an $\hat x$ polarized microwave field, $\Omega$, phase-locked to the $\hat y$ microwave.

The spin-glass phase can be prepared by starting in the paramagenetic phase and adiabatically decreasing $\Omega$ 
to some final value smaller than $V_{dd}$. If this is done slowly enough in a finite system, the quantum paramagnetic ground state adiabatically evolves across a second-order phase transition \cite{qsgreview} into the spin-glass ground state. 
In practice, the finite ramp speed will give rise to a finite density of defects (corresponding to an effective ``temperature'' that is proportional to the ramping rate) via the Kibble-Zurek mechanism~\cite{kibble, zurek2005}. However, due to the separation of timescales between $V_{dd}$ and the minimum ramp rate $\sim T_1^{-1}$, these defects should not obscure the visibility of the spin glass. In addition, this might amount to an experimentally efficient way of generating many-body localized excitations in the spin glass phase~\cite{mbl}.

The various realizable phases can be easily identified experimentally. The simplest case is the ferromagnet, which possesses net magnetization that can be  detected via spin-dependent fluorescence. Other phases, such as the paramagnet and the spin glass, have no net magnetization and thus cannot be distinguished via fluorescence. However, as shown in ref.~\cite{StrackPRL11}, there is a difference between their excitation spectra, which can be used to tell the paramagnet and the spin glass apart spectroscopically. Specifically, the linear response of the ground state to a $\hat y$-polarized rf field at various frequencies will give access to the spectral function according to Fermi's Golden Rule. In particular, deep in the gapped paramagnetic phase ($\Omega \gg V_{dd}$),  
there should be no rf absorption at frequencies below the transverse field $\Omega$, and a sharp peak at this frequency. As the transition into the spin glass is approached, this absorption peak moves to lower frequencies as the excitation gap of the paramagnet closes. 
The motion of the absorption peak to lower frequencies can be resolved provided that the scale of the gap is much larger than both $T_1^{-1}$ or $T_2^{-1}$; this condition holds in the regime of interest. In particular, as the transverse field $\Omega$ becomes comparable to the interaction strength $V_\text{dd}$, the peak is expected to approach zero frequency and remain there as a broad spectral feature for $\Omega < V_\text{dd}$. The presence of such a broad feature at zero frequency across a wide range of $\Omega$ (together with the absence of magnetic ordering) would strongly suggest the presence of a spin glass.
Furthermore, the distinctive dynamical properties of the spin glass --- in particular, its slow dynamics --- can also be studied using a variety of approaches such as quenches, which would probe aging and related phenomena~\cite{fisherhuse, AmirPNAS12}. 

Finally, we note that the proposed antiglass phase would be straightforward to identify, as it would correspond to a \emph{narrowing} of the absorption spectrum, or alternatively to well-defined Rabi oscillations~\cite{asg2002} at low transverse field.


\section{Applications to  Exotic Quantum Magnetism}
\label{sec:magnetism}

Powerful microwave dressing techniques borrowed from the polar-molecule literature \cite{micheli06,brennen07,buchler07,buchler07b,micheli07,gorshkov08c,lin10b,cooper09,wall09,wall10,schachenmayer10,kestner11,gorshkov11b,gorshkov11c,kuns11,manmana12,gorshkov13, LemeshkoPRL12,weimer13, LemeshkoReview13} can be used to access a great variety of Hamiltonians beyond the one given in Eq.~(\ref{Ising}). Fermionic isotopes ${}^{161}$Dy and ${}^{163}$Dy have nuclear spin $I=5/2$; the resulting rich hyperfine level structure paves the way to engineering a different and possibly wider range of Hamiltonians compared to what bosonic isotopes can offer. Following Refs.\ \cite{gorshkov11b,gorshkov11c,manmana12,gorshkov13}, microwave dressing can be used to isolate $(2 S + 1)$ dressed states  in each atom. Projecting dipole-dipole interactions on these states generates a Hamiltonian of the form $H = \frac{1}{2} \sum_{i \neq j} H_{ij}$, where \cite{gorshkov13}
\ba
R_{ij}^3 H_{ij} = \mathbf{v}(\theta_{ij},\phi_{ij}) \cdot \mathbf{H}, \label{eq:generalS}
\ea
with $\mathbf{v}(\theta,\phi) = (Y_{2,0}, \textrm{Re}[Y_{2,1}],\textrm{Im}[Y_{2,1}], \textrm{Re}[Y_{2,2}],\textrm{Im}[Y_{2,2}])$, where $Y_{2,m}(\theta,\phi)$ are the rank-2 spherical harmonics; $\mathbf{H}$ is a five-component vector of Hamiltonians acting on the Hilbert space of two spin-$S$ systems at sites $i$ and $j$. With a sufficient number of microwave fields, as well as with linear and quadratic Zeeman and Stark shifts (both AC and DC), it is possible to fully and independently control each of the five components of $\mathbf{H}$ subject only to the constraints of Hermiticity and symmetry under the exchange of $i$ and $j$. In the case of $S = 1/2$, dropping constant terms, Eq.\ (\ref{eq:generalS}) reduces to \cite{gorshkov13}
\ba
R_{ij}^3 H_{ij} &=& \mathbf{v}(\theta_{ij},\phi_{ij}) \cdot [\mathbf{B}_x (S^x_i + S^x_j)   \nonumber \\
&&+ \mathbf{B}_y (S^y_i + S^y_j) + \mathbf{B}_z (S^z_i + S^z_j)  \nonumber \\
&&+ \mathbf{J}_{xx} S^x_i S^x_j  + \mathbf{J}_{yy} S^y_i S^y_j+ \mathbf{J}_{zz} S^z_i S^z_j   \nonumber \\
&& + \mathbf{J}_{xy} (S^x_i S^y_j + S^y_i S^x_j) + \mathbf{J}_{xz} (S^x_i S^z_j + S^z_i S^x_j) \nonumber \\
&&+ \mathbf{J}_{yz} (S^y_i S^z_j + S^z_i S^y_j)],\label{eq:Sonehalf}
\ea
where $\mathbf{B}_x$, $\mathbf{B}_y$, $\mathbf{B}_z$, $\mathbf{J}_{xx}$, $\mathbf{J}_{yy}$, $\mathbf{J}_{zz}$, $\mathbf{J}_{xy}$, $\mathbf{J}_{xz}$, and $\mathbf{J}_{yz}$ are five-component vectors. 

In the case of electric dipole-dipole interactions between polar molecules, specific level configurations have already been obtained to realize a variety of special cases of Eqs.\ (\ref{eq:generalS},\ref{eq:Sonehalf}). In particular,  the $\mathbf{J}_{xy}$ and $\mathbf{J}_{zz}$ interactions of Eq.\ (\ref{eq:Sonehalf}) can be used to obtain an XXZ model with interactions featuring a direction-dependent spin-anisotropy. This model can be used to study symmetry-protected topological phases on a ladder \cite{manmana12}. Furthermore, the interactions $\mathbf{J}_{xx}$, $\mathbf{J}_{yy}$, and $\mathbf{J}_{zz}$ of Eq.\ (\ref{eq:Sonehalf}) can be used \cite{gorshkov13} to realize the Kitaev honeycomb model in  a nonzero magnetic field (realized using terms $\mathbf{B}_x$, $\mathbf{B}_y$, and $\mathbf{B}_z$) --- a model that supports exotic non-Abelian anyonic excitations \cite{kitaev06}. Finally, Eq.\ (\ref{eq:generalS}) can be used to realize the most general spin-1 bilinear-biquadratic model \cite{schollwock96} with the $Y_{2,0}$ angular dependence \cite{manmana12}. It would be particularly interesting to follow Refs.\ \cite{deng05, hauke10,peter12,koffel12,wall12,nebendahl12,cadarso12,manmana12} and study how all of these spin Hamiltonians and the underlying phase transitions are affected by the long-range nature of the interactions.

The specific examples mentioned above were developed in the context of highly ordered arrays of polar molecules trapped in optical lattices. One might be able to obtain arrays of Dy atoms in a solid matrix by using masks for implantation~\cite{VieuASC00}. However, the resulting positions of the atoms would still feature substantial uncertainty, which will enable experimental studies and motivate theoretical studies of the effects of disorder on the above mentioned and other exotic spin models.

At sub-wavelength distances, the readout of individual atoms can be performed using a number of techniques. One possibility is to rely on spectroscopic addressability \cite{thomas89} enabled by spatially varying magnetic fields \cite{stokes91,schrader04,derevianko04} or Stark shifts \cite{gardner93,zhang06,lee07}. Another option is to rely on the nonlinearity of the atomic response to light  and thus employ techniques such as STED \cite{hell07}, spin-RESOLFT \cite{johnson98,cho07,maurer10}, or dark-state-based techniques \cite{yavuz07,juzeliunas07,gorshkov08}.

\begin{table*}
\caption{\label{comparison} Comparison of the proposed platform  with the alternatives}
    \begin{tabular}{| l | l | l | l | l | }
    \hline
     System                   &  Spin number  &  Spacing (nm) &   $V_\text{dd}$ (Hz) &  $T_1$ (s) \\ \hline
    Matrix-isolated Dy (this work) & macroscopic                         & $10$           & $10^6$                             &  $\gtrsim 0.1$                           \\
        Solid-state (LiHoF)~\cite{qsgreview}               & macroscopic                         & $1$                   & $10^{9}$     & $<0.01$                        \\
            NV centers in diamond~\cite{JelezkoNJP12}             & macroscopic                         & $50$                  & $400$                      & $ \gtrsim 1$                              \\
                Dysprosium in optical lattices~\cite{LuPRL11}    & $10^5$                              & $500$                 & $10$                     & $ \gtrsim 1$             \\
    Polar molecules (KRb)~\cite{YanArxiv13}              & $10^4$                              & $1000$                 & $250$            & $ \gtrsim 1$                          \\
          \hline
    \end{tabular}
\end{table*}

\section{Conclusions}
\label{sec:conclusions}

In this work we presented a new platform for quantum simulation of many-body systems based on strongly magnetic atoms trapped in inert matrices, which can be complementary with other solid state approaches \cite{NormSteveNV}.  In particular, although solid-state defects such as NV centers in diamond feature naturally long room-temperature coherence times~\cite{BalasubramanianNatMat09}, their interactions are limited to tens of kHz at $\sim 10$~nm spacing~\cite{JelezkoNJP12, ChildressScience06, YaoNatComm12}.  In addition to stronger magnetic dipolar interactions, the density of magnetic atoms can, in principle, be controlled via direct changes during the embedding process; solid-state defects, on the other hand, often require high-temperature annealing steps which result in low conversion efficiencies.  On the other hand, compared to atoms in optical lattices~\cite{BlochRMP08}, the proposed scheme offers an additional feature of \textit{scalability}, which is particularly important to study the physics of disordered materials whose behavior is dominated by rare events. For reference purposes, we provide the comparison of matrix-isolated Dy with alternative platforms in Table~\ref{comparison}.

We have suggested that the platform can be used to study the properties of the spin-glass phase and reveal the existence of the antiglass phase in LiHo$_x$Y$_{1-x}$F$_4$, and discussed possible applications to exotic quantum magnetism. We focused on magnetic atoms isolated in solid helium in order to make use of the experimental data available on decoherence induced by a helium matrix. However, high impurity densities can be much easier achieved in matrices that are solid at ambient pressure, such as parahydrogen.\cite{MomoseVibSpec04} We hope that our paper will prompt the experimental studies in this direction. Furthermore, the proposed scheme can be realized with a variety of magnetic atoms, such as Cr or Er, and polar molecules, such as CO or OH.
Although in this work we have focused on coherent dynamics, the intrinsically driven nature of the proposed system allows to use it to study dissipative magnetic transitions~\cite{LeeArXiv13, BuchholdArxiv13}. Furthermore, the described approach is not limited to dipolar systems and can be applied to other types of interactions, e.g.\ quadrupole-quadrupole ones~\cite{KokshenevPRB96, BhongalePRL13}.


%

\section{acknowledgements}

We are indebted to Mikhail Lukin, Ariel Amir, and Benjamin Lev for insightful discussions. M.L. acknowledges support from NSF through a grant for the
Institute for Theoretical Atomic, Molecular, and Optical Physics at
Harvard University and Smithsonian Astrophysical Observatory. N.Y.Y. acknowledges support from DOE (FG02-97ER25308). A.V.G. acknowledges funding from the Lee A. DuBridge foundation and from the IQIM, an NSF Physics Frontier Center with support of the Moore foundation.  S.G. acknowledges support from the Harvard Quantum Optics Center. M.L. and A.V.G. thank KITP for hospitality.

\bibliography{References_library}

\end{document}